# Dust dissipation timescales in the intermediate and outer regions of protoplanetary disks

Hiroshi MAESHIMA[1,2,*], Takao NAKAGAWA,[1], Takuya KOJIMA[1,2], Satoshi TAKITA,[3] and Jungmi KWON[4]

[1]Institute of Space and Astronautical Science, Japan Aerospace Exploration Agency, 3-1-1 Yoshinodai, Chuo-ku, Sagamihara, Kanagawa 252-5210, Japan

[2]Department of Physics, Graduate School of Science, The University of Tokyo, 7-3-1 Hongo, Bunkyo-ku, Tokyo 113-0033, Japan

[3]National Astronomical Observatory of Japan, 2-21-1 Osawa, Mitaka, Tokyo 181-8588, Japan

[4]Department of Astronomy, Graduate School of Science, The University of Tokyo, 7-3-1 Hongo, Bunkyo-ku, Tokyo 113-0033, Japan

*E-mail: maeshima@ir.isas.jaxa.jp

**Abstract**

Dust and gas in protoplanetary disks dissipate as central stars evolve. In order to estimate the dust dissipation timescales in the protoplanetary disks, we stacked the WISE 12, 22, and the AKARI 90 $\mu$m survey images of known T Tauri stars and derived the average fluxes, well below the survey flux limit in the 90 $\mu$m band. We classified 4,783 T Tauri stars into three age groups, which are young (<2 Myr), mid-age (2–6 Myr), and old (>6 Myr) groups, and stacked the WISE 12 and 22 and the AKARI 90 $\mu$m images in each group. The photometry of the stacked image shows the flux decay timescales of 1.4±0.2, 1.38±0.05, and $1.4^{+0.6}_{-0.5}$ Myr in the 12, 22, and 90 $\mu$m bands, respectively. In optically thin disks with one-solar luminosity central stars, the 12 and 22 $\mu$m fluxes are attributed to the emission from the intermediate ($\sim$ 1 au) region and the 90 $\mu$m flux corresponds to that from the outer ($\sim$ 10 au) region in the disk. We hence conclude the dust dissipation timescale is $\tau_{\rm med,dust}$ $\sim$1.4 Myr in the intermediate disks and is $\tau_{\rm outer,dust}$ =$1.4^{+0.6}_{-0.5}$ Myr in the outer disks. The dust-dissipation time difference between the outer and intermediate disks is $\Delta\tau_{\rm dust} = \tau_{\rm outer,dust} - \tau_{\rm med,dust} = 0.0^{+0.6}_{-0.5}$ Myr, indicating that the dust in the intermediate and outer disks dissipates on almost the same timescale.

---

## 1 Introduction

Protoplanetary disks are circumstellar disks around pre-main-sequence stars, and they consist of gas and dust (e.g., Williams & Cieza 2011). Dust and gas, which are the raw materials of planetary systems, dissipate in disks as the central stars evolve. Therefore, studying the dissipation processes of dust and gas in protoplanetary disks is important for understanding planetary system formation.

The dissipation process is currently under discussion. In the viscous accretion model, the entire disk dissipates on the same timescale (e.g., Armitage 2010). On the other hand, the photoevaporation model predicts that the dissipation timescales differ between inner and outer disk regions (e.g., Gorti et al. 2009). By observing evidence of the dissipation timescales in multiple disk regions, we can restrict the process of disk clearing.

Observations in infrared (IR) wavelengths are effective in the study of dust dissipation timescale since the fluxes of disks at several IR wavelengths can inform the dust spatial distribution. Dust grains in a disk are heated by the central star and emit thermal radiation. Dust farther away from the central star is cooler and emits radiation that peaks at longer wavelengths. For example, the dust temperature in an optically thin disk around a star with one-solar luminosity is $T_{\text{dust}} \sim 280(r/1 \text{ au})^{-1/2}$ K at the distance $r$ from the central star (Wyatt 2008). Suppose that dust grains emit blackbody radiation, we can assume that the near-IR (a few $\mu$m), mid-IR ($\sim 10$ $\mu$m), and far-IR ($\sim 100$ $\mu$m) fluxes arrive mainly from the inner disks ($\sim 0.1$ au), intermediate disks ($\sim 1$ au), and outer disks ($\sim$10 au), respectively.

Dust dissipation timescales in the inner and intermediate disks were evaluated quantitatively by Ribas et al. (2014). They derived the decay timescales of the fraction of the IR excesses of young stellar objects (YSOs) in the near-IR and the mid-IR bands. The derived decay timescales tend to be longer at longer wavelength bands. The increase of decay timescales at longer wavelengths is compatible with inside-out disk clearing scenarios (e.g., Williams & Cieza 2011) at the inner and intermediate disks.

On the other hand, the timescale in the outer disks has not been observationally quantified. The outer-disk dust mainly emits radiations in far-IR, sub-mm, and mm wavelengths. To evaluate the decay timescale in the outer disks, we need to observe stars with evolved disks. T Tauri stars (TTSs) are low-mass pre-main-sequence (PMS) stars. Among them, weak-line T Tauri stars (WTTSs) have weak evidence of the accretion (e.g., Shu et al. 1987; Strom et al. 1989) and are considered to be at the final stages of disk clearing (Hardy et al. 2015). Hence we need to observe not only classical TTSs but also WTTSs. Atacama Large Millimeter/Submillimeter Array (ALMA) observed many disks with high sensitivity in sub-mm and mm wavelengths (e.g., Ansdell et al. 2016). However, the disks in the final evolutionary stage are relatively faint emitters at these wavelengths, so the disks in WTTSs are not easily detected even with ALMA (e.g., Hardy et al. 2015). Furthermore, the detector sensitivity of far-IR wavelengths is relatively low and the number of detected evolved disks is small. For example, Herschel and Spitzer detected only a few dozen WTTSs in the 70 $\mu$m band (Wahhaj et al. 2010; Cieza et al. 2013). Therefore, we cannot reliably discuss dust-dissipation timescales based on a small database of observations of individual sources. To obtain the typical dust-dissipation timescale in the outer disks, we should analyze the decay timescales of the far-IR or sub-mm fluxes of many disks, including faint disks that are classified as evolved disks.

In this paper, to enable reliable estimates of far-IR emission from TTSs, we carry out stacking analysis of TTSs in the mid-IR and far-IR bands and evaluate the dust dissipation timescales in the intermediate and outer disks. The stacking analysis is a method to observe average sources well below the flux limit by image stacking (e.g., Matsuki et al. 2017; Kojima et al. in preparation). Following the introduction in section 1, section 2 describes our sample-selection method for analysis and classification based on estimated ages. Section 3 describes the results in the mid-IR and far-IR bands. Based on these results, section 4 obtains the dust-dissipation timescales in the intermediate and outer disk regions. The summary is presented in section 5.

## 2 Method

This section describes our analysis method. First, we describe all-sky survey data to be used and the need for a stacking analysis in subsection 2.1. Subsection 2.2 describes the method of sample selection. Then, we estimate the ages of the samples and classify the samples into three age groups in subsection 2.3.

### 2.1 All-sky survey data

#### 2.1.1 AKARI and the need for stacking analysis

To collect many far-IR observations of TTSs, we use all-sky survey data in the far-IR wavelength taken by AKARI. AKARI is the Japanese infrared satellite launched in 2006 (Murakami et al. 2007). The all-sky survey with AKARI Far-Infrared Surveyor (FIS) covers more than 99% of the sky in the 65, 90, 140, and 160 $\mu$m bands (Kawada et al. 2007). This survey provides the far-IR observations of many objects, regardless of whether the objects can be detected or not. We use the AKARI all-sky survey data in the 90 $\mu$m band since the 90 $\mu$m band is the most sensitive for dust emission.

Although the noise level in the 90 $\mu$m band is the most sensitive among those in the FIS bands, the detector sensitivity is relatively lower than the typical flux of WTTSs. The 5$\sigma$ detection limit of the 90 $\mu$m band is 0.55 Jy (Arimatsu et al. 2014). On the other hand, we estimate the typical flux in the 90 $\mu$m wavelength as $\sim 0.01$ Jy based on the flux of WTTSs at 100–200 pc in a sub-mm wavelength (Hardy et al. 2015). Individual TTSs, especially WTTSs, are difficult to be detected with AKARI.

To solve the sensitivity problem, we perform a stacking analysis of the 90 $\mu$m images. The stacking analysis is an analysis method to obtain an average image with high signal-to-noise (*S/N*) ratio by averaging images of the same type of objects (e.g., Matsuki et al. 2017; Kojima et al. in preparation). We make the average image of TTS in the 90 $\mu$m band for known TTSs, trying to detect average flux well below the survey flux limit by improving the *S/N* ratios with the stacking analysis.

### 2.1.2 WISE

To compare the age dependences of the far-IR fluxes obtained by AKARI with those of the mid-IR fluxes, we use all-sky survey data by the Wide-field Infrared Survey Explorer (WISE). WISE is an infrared satellite launched in 2009 and performed all-sky surveys in the 3.4, 4.6, 12, and 22 $\mu$m bands (Wright et al. 2010). To compare the age dependences of the mid-IR fluxes with that of the far-IR fluxes, we use the survey data of AllWISE Image Atlas in the 12 and 22 $\mu$m bands (Cutri 2013).

## 2.2 Sample Selection

### 2.2.1 Selection of T Tauri Stars from a PMS star catalog

To discuss the dissipation timescale with a large number of samples, we make an input catalog of TTSs in the whole sky based on a PMS star catalog built by Zari et al. (2018). The PMS catalog is composed based on the second data release of the *Gaia* mission, *Gaia* DR2 (Collaboration et al. 2016; Collaboration et al. 2018). The catalog contains 43,719 PMS stars within $< 500$ pc selected based on the velocities and the relative positions to the PARSEC isochrone (Bressan et al. 2012) on the Hertzsprung-Russel (H-R) diagram.

Figure 1 shows the flowchart of the selection of TTSs for our stacking analysis and the remaining samples after each screening procedure. From the PMS catalog, we search PMS stars that have no other PMS stars within the radii of 90″( This procedure is shown as "a1" in figure 1). This screening is set to avoid sample confusion within the 90″ AKARI aperture (Takita et al. 2015).

There is a possibility of other bright infrared sources within the AKARI photometric aperture.

Since the AKARI aperture size is $90''$ and the full width at the half maximum (FWHM) of the point spread functions (PSF) is $\sim 60''$ (Takita et al. 2015), it is difficult to separate the target sources from contaminant sources from AKARI image. Therefore, we discard objects contaminated by nearby bright sources within the AKARI aperture based on the AllWISE Source Catalog (Cutri 2013) (a2). The AllWISE Source Catalog contains objects detected on the AllWISE Image Atlas. Since the FWHM in the 22 $\mu$m band is $\sim 12''$ (Cutri 2013), we can detect contaminant infrared bright sources within the $90''$ AKARI aperture. We need to discard the contaminant sources which can be detected in stacked images. The nominal $5\sigma$ detection limit in the 22 $\mu$m band is 6.7 mag, which is $F_{22\mu m,\mathrm{limit}} = 18$ mJy in flux (Wright et al. 2010). Then we assume that the contaminant object detected in the stacked image has the brightness $F_{22\mu m,\mathrm{bright}}$ satisfying the following inequality:

$$F_{22\mu m,\mathrm{bright}}/N_{\mathrm{stack}} > F_{22\mu m,\mathrm{limit}}/\sqrt{N_{\mathrm{stack}}},$$
$$F_{22\mu m,\mathrm{bright}} > F_{22\mu m,\mathrm{limit}}\sqrt{N_{\mathrm{stack}}}, \qquad (1)$$

where $N_{\mathrm{stack}}$ is the number of stacked images and the $F_{22\mu m,\ \mathrm{limit}}$ is the nominal $5\sigma$ detection limit. Here we assume the noise is dominated by the random noise and the detection limit of the stacked images is proportional to $1/\sqrt{N_{\mathrm{stack}}}$. When we stack $\sim 4,000$ images, the brightness limit is needed to be $F_{22\mu m,\mathrm{bright}} > 1.1$ Jy. Thus, we define the contaminant sources as sources with $> 1.1$ Jy in the 22 $\mu$m band. To exclude samples with the contamination, we discard the samples which have $\geq 1$ contaminant sources in the annulus range of $16.''5$–$90''$ or $\geq 2$ contaminant sources in the radius of $16.''5$ around the coordinates. Here, the contaminant search ranges are determined based on the photometric apertures ($16.''5$ and $90''$ in the 22 and 90 $\mu$m bands, respectively; Cutri 2013; Takita et al. 2015). After (a2) screening, 40,304 samples remain.

If we include distant sources in the stacking analysis, the *S/N* ratio will decrease due to the reduction of the average intensity. Thus, we must limit the distance range of the samples. Among samples chosen by (a2) screening, we select samples whose distances are 100–200 pc (a3) since most of the nearby young clusters are at 100–200 pc (e.g., Ribas et al. 2014). After (a3) screening, 6,494 samples remain.

Then, we should restrict the range of the stellar mass to reduce the bias due to stellar mass. We limit the sample to low-mass stars, by which the number of remaining samples is dominated. we can roughly limit the stellar-mass range by limiting the intrinsic color. A K0-type main-sequence star has the effective temperature $T_{\mathrm{eff}} \sim 5250$ K (Kenyon & Hartmann 1995). From the color-temperature relation studied by Mucciarelli & Bellazzini (2020), stars with $T_{\mathrm{eff}} \sim 5250$ K are estimated to have the color of $G_{BP} - G_{RP} = 0.994$ mag. Then, we limit the samples with the intrinsic color of $(G_{BP} - G_{RP})_0 > 0.994$ mag to restrict the range of the stellar mass (a4). The intrinsic color is derived as

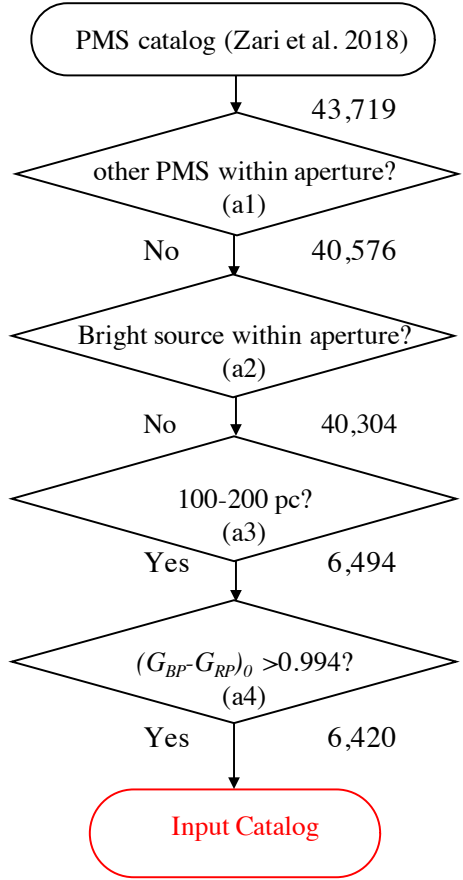

**Fig. 1.** Flowchart of sample selection for the input catalog. The sample selection screenings are described in sub-subsection 2.2.1. The number at the bottom right of each box indicates the number of samples after each screening. (Color online)

$$(G_{BP} - G_{RP})_0 = G_{BP} - G_{RP} - E(G_{BP} - G_{RP}), \tag{2}$$

where $E(G_{BP} - G_{RP})$ is the color excess derived by Zari et al. (2018).

After (a1)–(a4) screening, we finally select 6,420 TTSs from the PMS catalog, and compile them into the input catalog for our analysis.

#### 2.2.2 AKARI image check

For every source in the input catalog described above, we check each AKARI 90 $\mu$m image to see if it is suitable for the stacking analysis. We use the Far-Infrared All-Sky Survey Maps public release ver 1 survey (Doi et al. 2015). In this released data, the all-sky image is divided into $6° \times 6°$ image sets. From the $6° \times 6°$ images of the 90 $\mu$m all-sky survey, we make $10'.25 \times 10'.25$ images whose central positions coincide with the coordinates of samples in the input catalog.

Figure 2 shows the flowchart of the AKARI image check. We require the images of the objects to satisfy the following five criteria for the 90 $\mu$m stacking analysis:

(b1): We select $10'.25 \times 10'.25$ cutout images that do not protrude from the original $6° \times 6°$ map.

(b2): We select images in which the scan number $N_{\rm scan}$ of the pixels is $N_{\rm scan} \geq 2$ because the intensity values in regions with $N_{\rm scan} < 2$ are unreliable (Matsuki et al. 2017).

(b3): We select images that do not contain $\geq 10$ pixels with values below $-10$ MJy sr$^{-1}$. This is because such pixels are probably affected by high-energy particles (Doi et al. 2015).

(b4): We select images that do not contain $3 \times 3$ pixel areas with total pixel values above 200 MJy sr$^{-1}$ located at $\geq 90''$ from the image center because such areas are assumed to be contaminated by nearby bright sources (Matsuki et al. 2017).

(b5): Images with high standard deviations of the sky are too noisy for stacking analysis. Thus, we select images in which the standard deviation $\sigma_{\rm sky}$ of the area with radii of 120–300$''$ is below $5\Delta\sigma_{\rm sky} = 6.62$ MJy sr$^{-1}$. Here, the $\Delta\sigma_{\rm sky}$ is the standard deviation of the $\sigma_{\rm sky}$ values of our samples surviving the (b4) screening operation.

Through screening (b1)–(b5), we select 4,808 objects from the input catalog for stacking analysis in the 90 $\mu$m band.

## 2.3 Age estimation and age categorization of the samples

To evaluate the age dependence of the flux in the mid-IR and far-IR wavelengths, we estimate the ages of individual objects selected in subsection 2.2. Here, details of the age estimation are described. The ages of samples are estimated by comparing the positions of samples on the H-R diagram and an isochrone model.

To estimate age with the isochrone model, we need the information of the intrinsic color $(G_{BP} - G_{RP})_0$ and the $G$-band absolute magnitude $M_G$ for individual objects. The absolute magnitude $M_G$ is derived as follows:

$$M_G = G - A_G + 5(\log_{10}(\varpi) + 1), \tag{3}$$

Here, $G$ is the $G$-band apparent magnitude, $A_G$ is the extinction in the $G$-band, and $\varpi$ is the parallax. The values of $G$, $A_G$, $\varpi$ are taken from the PMS catalog (Zari et al. 2018). The intrinsic color $(G_{BP} - G_{RP})_0$ is derived in the same way described in sub-subsection 2.2.1.

Then, we estimate the age of individual objects by comparing objects and an isochrone model on the H-R diagram. We use the PARSEC isochrone (Bressan et al. 2012) version 1.2S (Chen et al. 2014; Tang et al. 2014). We set the model with $A_V = 0$ mag and solar metallicity $Z = 0.0152$. The photometric systems of *Gaia*'s DR2 $G$, $G_{BP}$, $G_{RP}$ bands (Evans et al. 2018) are used in the model. We estimate ages as the values linearly-interpolated from the isochrone grids. Figure 3 shows the position of the samples and the PARSEC isochrone on the H-R diagram. There are 25 objects with $(G_{BP} - G_{RP})_0 > 3.8$ mag, where the data table of the PARSEC isochrone model is not constructed.

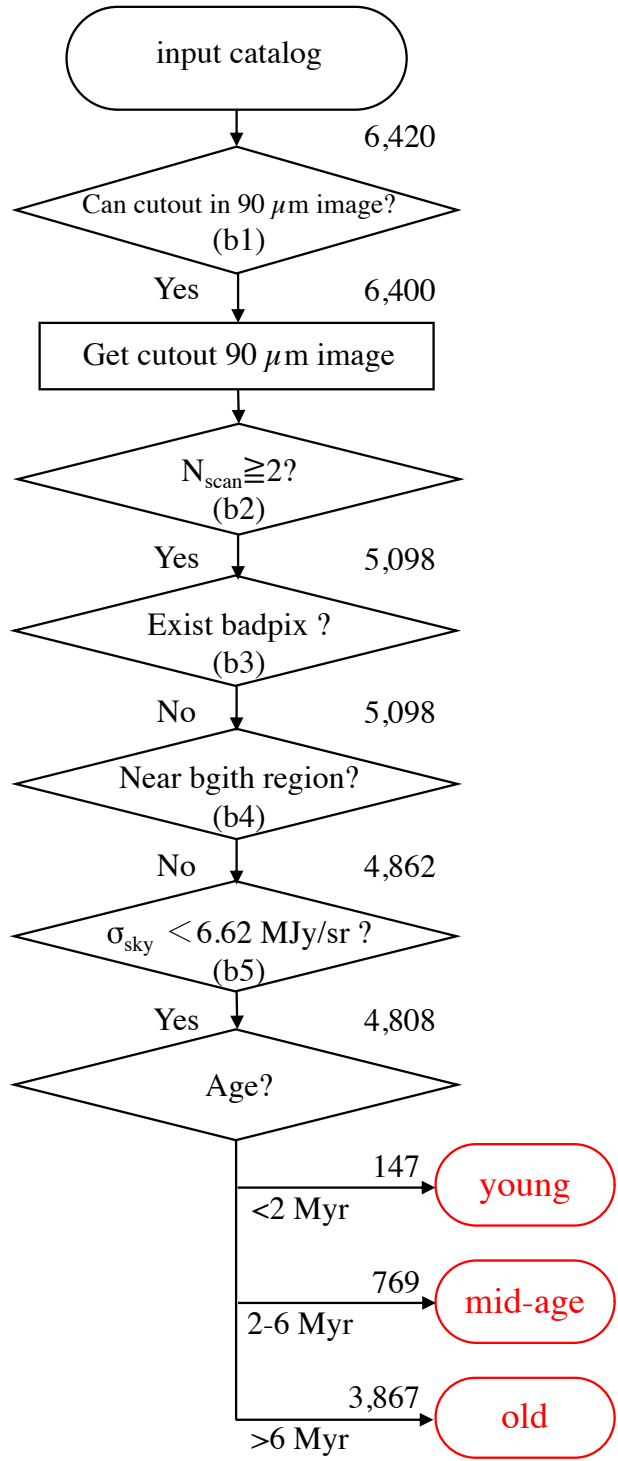

**Fig. 2.** Flowchart of selecting suitable objects for the 90 $\mu$m stacking analysis and their classification by age. The number on the side of arrows indicates the number of samples after each screening or classification. The selection criteria and the age classification are described in sub-subsection 2.2.2 and subsection 2.3, respectively. (Color online)

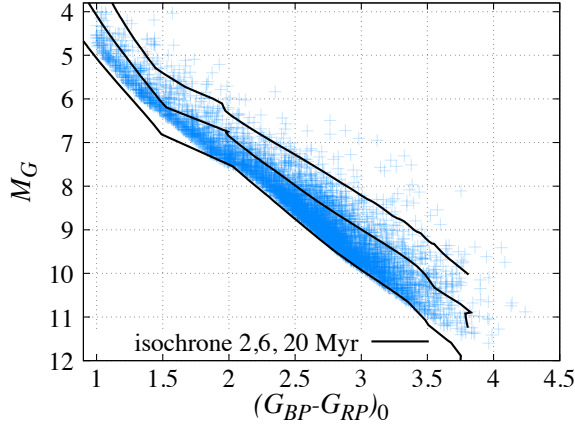

**Fig. 3.** The locations of the samples on the H-R diagram. The vertical axis shows the absolute $G$- band magnitude $M_G$, and the horizontal axis shows the intrinsic color $(G_{BP} - G_{RP})_0$. The blue points show the individual samples. The lines show the PARSEC isochrone line at 2, 6, 20 Myr from top to bottom. There is a lack of objects with $(G_{BP} - G_{RP})_0 < 1.8$ between 6 Myr and 20 Myr isochrone lines because Zari et al. (2018) discard binary sequence star candidates, which are brighter than the main-sequence by 0.75 mag. (Color online)

Therefore, we discard the samples with $(G_{BP} - G_{RP})_0 > 3.8$ mag. Figure 4 shows the histogram of the estimated ages of samples. The estimated ages are less than $\sim 20$ Myr since Zari et al. (2018) limit samples with age $< 20$ Myr in the PMS catalog.

To make average images of TTSs of similar age, we divide the samples into three age groups: younger than 2 Myr (the young group), between 2 Myr and 6 Myr (the mid-age group), and older than 6 Myr (the old group). The young, mid-age, and old groups contain 147, 769, and 3,867 objects, respectively. The typical age is defined as the median age of the samples in the group and those of the young, mid-age, and old groups are derived to be $1.23^{+0.03}_{-0.04}$, $4.44^{+0.03}_{-0.04}$, and $11.96^{+0.05}_{-0.04}$ Myr, respectively. Here, the uncertainty of the typical age is estimated to be $+(Q_3 - Q_2)N_{\mathrm{stack}}^{-1/2}$, $-(Q_2 - Q_1)N_{\mathrm{stack}}^{-1/2}$, where $Q_i$ is the $i$-th quantile of the age distribution in the analyzed group and $N_{\mathrm{stack}}$ is the number of samples in each group.

From the methods described in subsections 2.2 and 2.3, we obtain three age groups of TTSs for stacking analysis in the AKARI 90 $\mu$m band. By performing the stacking analysis, we expect to obtain an average 90 $\mu$m image for each age group. We use this age classification not only for the stacking analysis with the AKARI 90 $\mu$m band but also for those with the WISE 12 and 22 $\mu$m bands to compare stacking-analysis results in different wavelengths with the same TTS samples (see subsection 3.2).

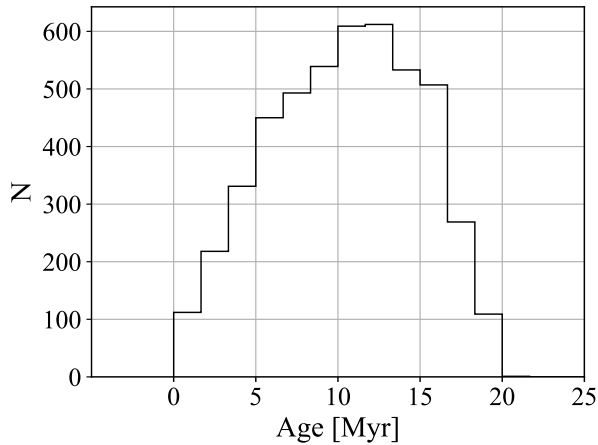

**Fig. 4.** Age histogram of the samples after (b5) screening. Twenty-five samples whose ages cannot be estimated are excluded. The number of samples in each bin is denoted by $N$.

## 3 Stacking Analysis and Results

This section describes obtained results. We first describe the stacking analysis results in the $90\ \mu$m band in subsection 3.1. Next, we describe the results in the $12$ and $22\ \mu$m bands in subsection 3.2. Then, we compare the age dependences of fluxes in the $12$, $22$, and $90\ \mu$m bands in subsection 3.3.

### 3.1 Stacking analysis in the 90 $\mu$m band

#### 3.1.1 Stacking analysis of each age group

For each age group, we perform the stacking analysis of the $90\ \mu$m images. Each image is stacked with the same weight. Figure 5 shows a single TTS image and the stacked images of the young, mid-age, and old age groups. As shown in figure 5, the stacked images have less noise than the single TTS image.

Table 1 shows the photometric result of the stacked images in the $90\ \mu$m band. The photometry in the $90\ \mu$m band is performed with Takita et al. (2015) method, with a $90''$ aperture radius and a sky-background annulus area of the radial range of $120$–$300''$. In the flux calibration, we assume the observed-to-expected flux ratio as $0.696$ (Takita et al. 2015).

We estimate the uncertainty of the fluxes as the standard deviation of photometric values of stacked sources of the randomly selected areas (see sub-subsection 3.1.2). In the young group, the stacked point source is detected at the $3\sigma$ level. On the other hand, the point source of the mid-age group is detected at the $1\sigma$ level and that of the old group is not detected. Comparing these three stacked images, we confirm that the $90\ \mu$m average fluxes tend to be fainter in older groups than their younger counterparts.

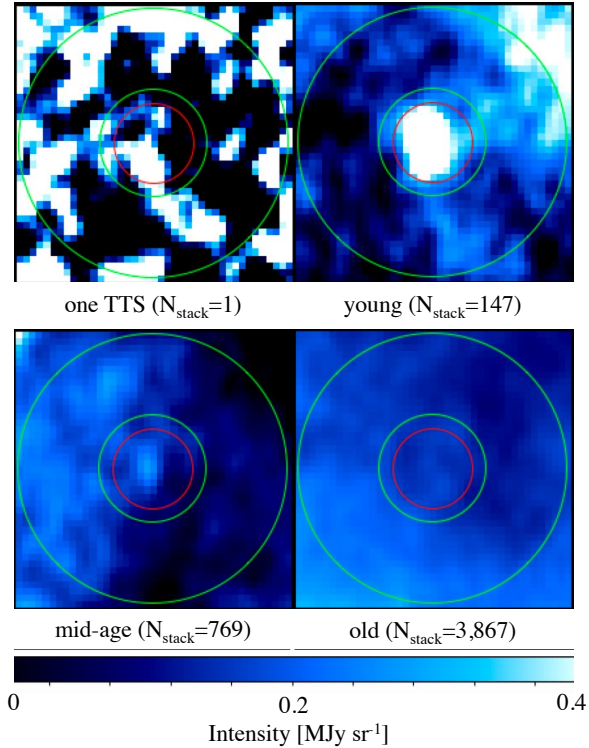

**Fig. 5.** A single TTS image (upper left) and the stacked images of the young (upper right), mid-age (lower left) and old (lower right) groups, respectively. The red circles delineate the $90''$ radius aperture. The regions enclosed by the two green circles are the sky annuli of the $120''$–$300''$ radial range. The field of view is $10'.25 \times 10'.25$. The color bar is the intensity scale in the 90 $\mu$m band. (Color online)

Here each image is stacked in the same weight since the effect of distance on brightness is assumed to be negligible due to the distance of 100–200 pc restricted by step (a3) in sub-subsection 2.2.1. Also, we assume that the stellar color dependence of the infrared intensity is negligible since we restrict the stellar intrinsic color $(G_{BP} - G_{RP})_0$. Possible effects of the distance and color correction on the stacking analysis are discussed in appendix 1.

### 3.1.2 Noise reduction by stacking in the AKARI 90 $\mu$m band

In this sub-subsection, we check whether the stacking analysis method in the AKARI 90 $\mu$m band reduces the noise level. If random noise dominates the total noise, the flux uncertainty of the stacked sources should be proportional to $N_{\text{stack}}^{-1/2}$, where $N_{\text{stack}}$ is the number of samples of each group. We perform the analysis of stacking images of the randomly selected areas from the regions near samples, called the random-stacking analysis.

In the following, we describe the procedure of the random-stacking analysis. Operations (r1)–(r4) are repeated until $N_{\text{stack}}$ cutout images are obtained:

(r1): To ensure that the sky condition is similar to that of the samples stacked in subsection 3.1, we need to select the area of random stacking near the samples. Thus, we randomly select one sample from the young, mid-age, or old group.

**Table 1.** Results of stacking analysis in the 12, 22, and 90 $\mu$m bands: the number of stacked samples $N_{\rm stack}$, the typical age, and the photometric fluxes of the stacked sources $F_{{\rm stacked},\nu}$. The effective wavelength is denoted by $\lambda_{\rm eff}$. The error range of the $F_{{\rm stacked},\nu}$ shows $1\sigma$ uncertainty range.

| | | young | mid-age | old |
|---|---|---|---|---|
| | $N_{\rm stack}$ | 147 | 769 | 3,867 |
| | age [Myr] | $1.23^{+0.03}_{-0.04}$ | $4.44^{+0.03}_{-0.04}$ | $11.96^{+0.05}_{-0.04}$ |
| band | $\lambda_{\rm eff}$ [$\mu$m] | | $F_{{\rm stacked},\nu}$ [mJy] | |
| W3 | 12 | $125.90 \pm 0.06^*$ | $18.93 \pm 0.03^\dagger$ | $5.00 \pm 0.01$ $^\ddagger$ |
| W4 | 22 | $246.8 \pm 0.2$ | $25.9 \pm 0.1$ | $3.63 \pm 0.07^\ddagger$ |
| WIDE-S | 90 | $327 \pm 51$ | $33 \pm 22$ | $< 30\ (3\sigma)$ |

$^*$There are 146 images stacked after discarding one image in the insufficient coverage area.
$^\dagger$There are 767 images stacked after discarding two images in the insufficient coverage area.
$^\ddagger$There are 3,865 images stacked after discarding two images in the insufficient coverage area.

(r2): We randomly select an off-center coordinate within a $\pm 0°\!.5 \times \pm 0°\!.5$ region from the coordinate of the sample selected in step (r1).

(r3): From the AKARI 90 $\mu$m all-sky survey, we make a $10'\!.25 \times 10'\!.25$ image whose central coordinate is selected in step (r2).

(r4): The screening condition for random-stacking analysis should be the same as the selection in subsection 2.2 as far as possible. Thus, we apply criteria (b1)–(b5) described in sub-subsection 2.2.2 to the image made in step (r3). We preserve the image if it meets all the criteria. If the image violates one or more criteria, we discard the image and return to step (r1).

By repeating the above operation (r1)–(r4), we obtain the $N_{\rm stack}$ images. Then, we stack them and perform the aperture photometry to the stacked sources. The stacking and photometric methods are the same as those described in subsection 3.1. In the case of $N_{\rm stack} = 1, 10, 10^2, 10^3$, and $5 \times 10^3$, we repeat the random stacking $N_{\rm repeat} = 10^4, 10^4, 10^3, 10^2$, and 30 times, respectively. We derive the standard deviation $\sigma_{\rm random}$ of the photometric values of the randomly stacked images.

The uncertainties $\Delta \sigma_{\rm random}$ of the $\sigma_{\rm random}$ are derived from the variance of the unbiased vari-

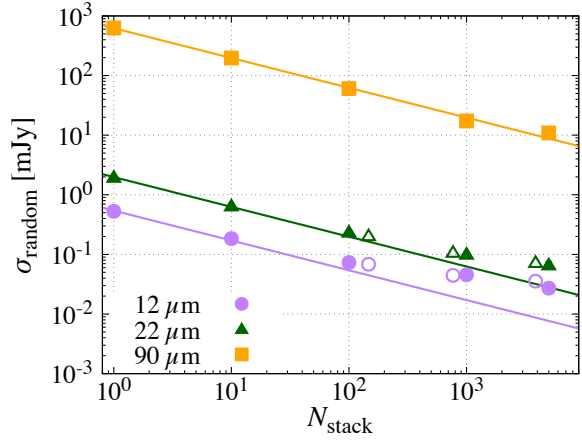

**Fig. 6.** The result of the random-stacking analysis in the 12 $\mu$m (purple circles), 22 $\mu$m (dark-green triangles), and 90 $\mu$m bands (orange squares). Open symbols show the $\sigma_{\rm random}$ for the $N_{\rm stack} = 147,\ 769$ and $3{,}867$ cases in the 12 and 22 $\mu$m bands. The details are described in sub-subsections 3.1.2 and 3.2.2. Each line shows a fitted function, $\sigma_{\rm random} = \sigma_0 N_{\rm stack}^{-1/2}$, with a free parameter $\sigma_0$ in each bands. The uncertainties of the $\sigma_{\rm random}$ are $\Delta\sigma_{\rm random} = 1.6\text{--}4.1 \times 10^{-3}$ mJy, $0.5\text{--}1.4 \times 10^{-2}$ mJy and $1\text{--}5$ mJy for the 12, 22, and 90 $\mu$m bands case, respectively. (Color online)

ance $(\Delta(\sigma_{\rm random}{}^2))^2$ for normal distribution as follows:

$$(\Delta(\sigma_{\rm random}{}^2))^2 = \frac{2\sigma_{\rm random}{}^4}{N_{\rm repeat} - 1}, \tag{4}$$

$$\Delta\sigma_{\rm random} = \frac{\sqrt{(\Delta(\sigma_{\rm random}{}^2))^2}}{2\sigma_{\rm random}}. \tag{5}$$

The derived uncertainty is $\Delta\sigma_{\rm random} = 1\text{--}5$ mJy for each $N_{\rm stack}$ case.

Figure 6 shows the relation between the $N_{\rm stack}$ and the standard deviation $\sigma_{\rm random}$ of the photometric values of the randomly stacked images. After least-square fitting, the relation between $\sigma_{\rm random}$ and $N_{\rm stack}$ is obtained as

$$\begin{aligned}\sigma_{\rm random, 90\mu m}(N_{\rm stack}) &= \sigma_0 N_{\rm stack}{}^{-1/2} \\ &= (623 \pm 4) N_{\rm stack}{}^{-1/2}\ [{\rm mJy}],\end{aligned} \tag{6}$$

where the free parameter is $\sigma_0$. As shown in figure 6, the $\sigma_{\rm random}$ decreases proportionally to $N_{\rm stack}{}^{-1/2}$. Therefore, in the 90 $\mu$m band, we conclude that the background noise is dominated by random noise and the stacking analysis is effective in reducing the noise level.

Since the expected value of the random-stacking photometry should be zero, the $\sigma_{\rm random}$ is a scatter of the photometric values with the same expected value. Thus, we assume uncertainties of the flux of the stacked object as the $\sigma_{\rm random, 90\mu m}(N_{\rm stack})$.

## 3.2 Stacking analysis in the 12 and 22 $\mu$m bands

### 3.2.1 Stacking analysis of each age group

To compare the age dependences of the far-IR fluxes obtained by AKARI (subsection 3.1) with those of the mid-IR fluxes, we perform stacking analysis for mid-IR images taken by WISE.

In this subsection, we describe the stacking analysis of the mid-IR bands. The samples and groups in this analysis are the same as those stacked in the 90 $\mu$m band. From the WISE all-sky survey maps in the 12 and 22 $\mu$m bands, we make $2'.475 \times 2'.475$ images whose central positions are the coordinates of the samples in each group. This cutout size is determined based on the size needed for the aperture photometry (Cutri 2013). Then, we stack the cutout images in each band for each group. However, the young group includes one sample within the zero-coverage areas in the 12 $\mu$m band, the mid-age group includes two zero-coverage samples in the 12 $\mu$m band, and the old group includes two zero-coverage samples in the 12 and 22 $\mu$m bands. Therefore, we discard these samples of each group when stacking in each band.

The 12 and 22 $\mu$m stacked images are subjected to the photometric procedure in "Explanatory Supplement TOC II. User's Guide to the WISE Preliminary Data Release" (Cutri 2013). The photometry of the 12 $\mu$m stacked images is performed with an $8''.25$ aperture radius and sky-background annulus area of the $50''$–$70''$ radial range. The photometry for the 22 $\mu$m case is performed with a $16''.5$ aperture radius and the sky annulus of $50''$–$70''$. The flux is calibrated by the observed-to-expected flux ratios, which are 1/1.84 and 1/1.76 in the 12 and 22 $\mu$m bands, respectively (Cutri 2013). We estimate the uncertainties of the fluxes from the standard deviation of photometric values of stacked images of randomly selected areas (see sub-subsection 3.2.2).

Table 1 contains the results of the stacking analysis in the 12 and 22 $\mu$m bands. As shown in table 1, All the stacked point sources are detected above the $3\sigma$ level in the 12 and 22 $\mu$m bands.

### 3.2.2 Noise reduction by stacking analysis in the WISE bands

To evaluate the uncertainty of the flux of the stacked images, we perform the random-stacking analysis in the 12 and 22 $\mu$m bands.

In this subsection, we describe the procedure of the random-stacking analysis for the 12 and 22 $\mu$m band cases. The main difference between the procedure for this case and that for the 90 $\mu$m band case (sub-subsection 3.1.2) is the treatment of outliers due to the high frequency of the contamination of the bright point source (r4' in the following). The following operations (r1')–(r4') are repeated until $N_{\mathrm{stack}}$ cutout images are obtained:

(r1'): We randomly select one sample from the young, mid-age, or old groups.
(r2'): We randomly select an off-center coordinate within a $\pm 0°.5 \times \pm 0°.5$ region from the coordinate of the sample selected in step (r1').

(r3'): We make a 2′.475×2′.475 image whose central coordinate is selected in (r2').

(r4'): We perform the aperture photometry to the image clipped in (r3'). To reduce the effect of outliers of the photometric values, we set the criterion value of $5\sigma_{\text{random},0} = 2.2$ and 8.3 mJy in the 12 and 22 $\mu$m, respectively, where $\sigma_{\text{random},0}$ is the standard deviation of the photometric values of the image randomly clipped by (r1')–(r3') screenings. If the absolute photometric value is $\leq 5\sigma_{\text{random},0}$, we preserve the image. If this value exceeds $5\sigma_{\text{random},0}$, we discard the image to remove outliers and return to (r1'). WISE detected more point sources in the 12 and 22 $\mu$m bands than those AKARI detected in the 90 $\mu$m band (Cutri 2013; Yamamura et al. 2009). If the random areas contain point sources, the photometric result of the images will show high fluxes and these outliers will occur. To evaluate the sky $S/N$ ratio, we should remove these outliers by (r4') screening.

Operations (r1')–(r4') are repeated until $N_{\text{stack}}$ cutout images are obtained. After that, we average the $N_{\text{stack}}$ images and perform the photometry to the stacked image. We repeat the above random stacking $N_{\text{repeat}} = 10^4$, $10^3$, $10^3$, $10^2$, and 30 times for the $N_{\text{stack}} = 1$, 10, $10^2$, $10^3$, and $5 \times 10^3$ cases, respectively. Then, we derive the standard deviation $\sigma_{\text{random}}$ of the photometric values for each $N_{\text{stack}}$ case.

Figure 6 contains the relation between $N_{\text{stack}}$ and the $\sigma_{\text{random}}$ in the 12 and 22 $\mu$m bands. The uncertainties of the standard deviation are $\Delta\sigma_{\text{random}} = 1.6$–$4.1\times 10^{-3}$ mJy and $0.5$–$1.4\times 10^{-2}$ mJy in the 12 and 22 $\mu$m bands, respectively. The derivation methods of these uncertainties are the same way as those for the 90 $\mu$m band case (see sub-subsection 3.1.2). From the results in the 12 and 22 $\mu$m bands, the stacking analysis reduces the noise level, although the $\sigma_{\text{random}}$ is not proportional to $N_{\text{stack}}^{-1/2}$ especially for the $N_{\text{stack}} \geq 100$ case.

We assume uncertainties of the flux of the stacked object as the $\sigma_{\text{random}}(N_{\text{stack}})$ with the same number of stacks by repeating the random stacking with $N_{\text{repeat}} = 1000$, 300, and 100 for the young, mid-age, and old groups, respectively.

3.3 Age dependence of the disk fluxes

As described in section 1, the fluxes in different IR wavelengths are considered to come from different disk regions. To compare the evolution of each disk region, we derive the age dependence of the fluxes in each wavelength band.

First, to identify the main sources of the fluxes, we estimate the emission from the photosphere. Especially for classical TTSs, the emission from the boundary layer and the disk will contaminate the photospheric emission at $\lesssim 0.8$ $\mu$m and $\gtrsim 2$ $\mu$m (Kenyon & Hartmann 1990). Also, the extinction is severer at shorter wavelengths than that at longer ones. Therefore, we estimate the photospheric flux using the photometric values in the $J$ band (1.235 $\mu$m), $H$ band (1.662 $\mu$m), and $K_s$ band (2.159 $\mu$m). The fluxes of the individual samples in the $J$, $H$, $K_s$ bands are derived by positional cross-match with the Two-Micron All-Sky Survey (2MASS) Point Source Catalog (Skrutskie et al. 2006). The extinction ratio $A_G/A_V$ is derived using the extinction curve with $R_V \equiv A_V/E(B-V) = 3.1$ (O'Donnell et al. 1994; Cardelli et al. 1989), and the $A_X/A_V (X = J, H, K_s)$ is taken from "The Asiago Database on Photometric Systems" (Fiorucci & Munari 2003). The extinction ratios are derived as

$$A_J/A_G = 0.32, \quad A_H/A_G = 0.20, \quad A_{K_s}/A_G = 0.14. \tag{7}$$

The de-reddened photometric fluxes of the stacked sources in the $J$, $H$, $K_s$ bands are determined by averaging individual fluxes in each group.

Then, we compare the emission from the photosphere and the obtained average fluxes. Figure 7 shows the spectral energy distributions (SEDs) of the stacked objects and those of the estimated photospheres. We extrapolate the photospheric fluxes from the de-reddened photometric fluxes in the $J$, $H$, and $K_s$ bands, assuming a black body photosphere. As shown in figure 7, the stacked sources of the young and mid-age group have IR excesses in the 12, 22, and 90 $\mu$m bands. On the other hand, the stacked source of the old group has IR excess in the 22 $\mu$m band, but we cannot see the excess in the 12 and 90 $\mu$m bands.

To discuss the disk evolution, we need to separate the flux derived from disks and that from photospheres. Then, we define the disk flux $F_{\text{disk},\nu}$ as

$$F_{\text{disk},\nu} = F_{\text{stacked},\nu} - F_{\text{photosphere},\nu}, \tag{8}$$

where $F_{\text{photosphere},\nu}$ is the extrapolated photospheric flux. Table 2 shows the $F_{\text{disk},\nu}$ of each group in each band. The uncertainty of the $F_{\text{disk},\nu}$ is derived by the root sum square of the uncertainty of the $F_{\text{stacked},\nu}$ and $F_{\text{photosphere},\nu}$. The uncertainty of the $F_{\text{photosphere},\nu}$ is estimated by the propagation of the fitting parameter errors.

Figure 8 shows the relation between the age and the disk flux $F_{\text{disk},\nu}$ in each band. The young and mid-age samples in the 12, 22, and 90 $\mu$m bands are fitted to the following exponential functions:

$$F_{\text{disk},\nu} = F_{0,\nu} \exp(-t/\tau_\nu), \tag{9}$$

where $F_{0,\nu}$ and $\tau_\nu$ are free parameters. Table 3 shows the results of the fitted parameters, $F_{0,\nu}$ and $\tau_\nu$, in each band. We can determine the parameters only with two data points of the young and mid-age groups for the 12 and 90 $\mu$m band cases due to non-detection of the $F_{\text{disk},\nu}$ of the old group. Therefore, the above curve-fittings exclude the plotted points of the old group not only in the 12 $\mu$m- and 90 $\mu$m-band but also in the 22 $\mu$m band to compare the three decay timescales within the same age range.

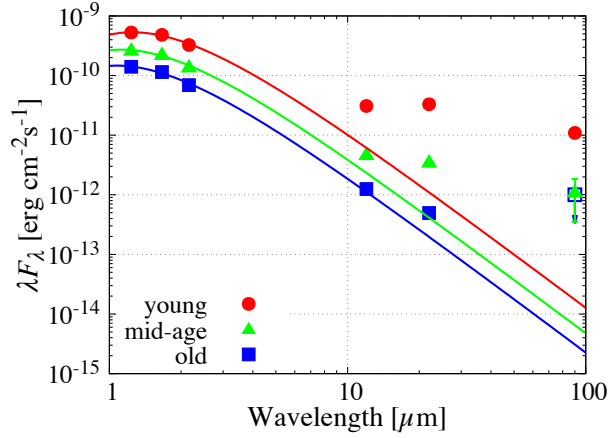

**Fig. 7.** Spectral energy distributions of the young (red circles) mid-age (green triangles) and old (blue squares) sample groups. Open symbols show the $3\sigma$ upper limit of the flux. The symbols in the 1.235, 1.662, and 2.159 $\mu$m bands are the mean de-reddened flux in the $J$, $H$, and $K_S$ bands, respectively. The lines show the estimated photospheric fluxes, assuming a single blackbody flux. (Color online)

**Table 2.** The disk flux $F_{\mathrm{disk},\nu}$ of each group in each band. The error range shows $1\sigma$ uncertainty range.

| $\lambda_{\mathrm{eff}}$ [$\mu$m] | young | mid-age | old |
|---|---|---|---|
|  | $F_{\mathrm{disk},\nu}$ [mJy] | | |
| 12 | $100 \pm 4$ | $10 \pm 2$ | $< 3\,(3\sigma)$ |
| 22 | $234 \pm 1$ | $22.8 \pm 0.7$ | $2.3 \pm 0.4$ |
| 90 | $326 \pm 51$ | $33 \pm 22$ | $< 30\,(3\sigma)$ |

The uncertainties of the parameters are derived by the error propagation law from the uncertainties of the flux and age. The decay timescales of the fluxes in the 12, 22, and 90 $\mu$m bands are consistent within the $1\sigma$ uncertainty range with each other. The discrepancy between the fitted curve and data of the old group in the 22 $\mu$m band is discussed in appendix 2.

**Table 3.** The fitted initial disk flux $F_{0,\nu}$ and decay time $\tau_\nu$ in each band. The error range shows $1\sigma$ uncertainty range.

| $\lambda_{\mathrm{eff}}$ [$\mu$m] | $F_{0,\nu}$ [mJy] | $\tau_\nu$ [Myr] |
|---|---|---|
| 12 | $240^{+30}_{-40}$ | $1.4 \pm 0.2$ |
| 22 | $570^{+60}_{-70}$ | $1.38 \pm 0.05$ |
| 90 | $800^{+200}_{-500}$ | $1.4^{+0.6}_{-0.5}$ |

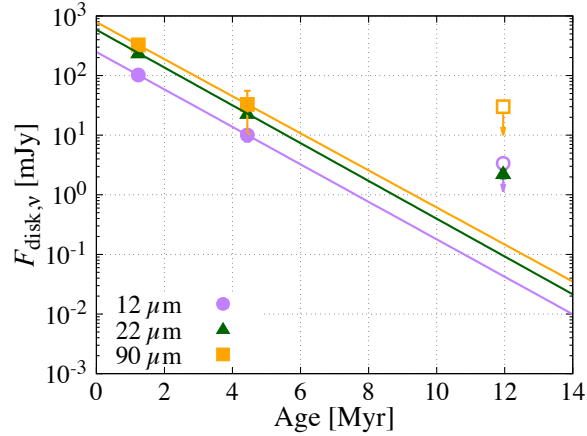

**Fig. 8.** Relation between the disk flux and age in each band: the 12 (purple squares), 22 (dark-green circles) and 90 (orange triangles) $\mu$m bands. Open symbols show the 3$\sigma$ upper limit of the $F_{\mathrm{disk},\nu}$. Lines are the exponential fittings to the young and mid-age disk fluxes in each band. (Color online)

## 4 Discussion

This section describes discussions based on the result derived in section 3. In subsection 4.1, we estimate the dust dissipation timescale in the outer disks based on the decay timescale of the 90 $\mu$m flux and compare our results with those of previous observational studies of the outer disk dust masses. In subsection 4.2, we compare the decay timescales at the far-IR and mid-IR bands and discuss the dust dissipation timescale in the intermediate and outer disks. Next, we compare our result with the decay timescales of debris disks, which are more evolved disks than protoplanetary disks, in subsection 4.3.

### 4.1 Estimation of dust mass and comparison with other studies of the outer disks

To discuss the dust dissipation timescale, we estimate the dust mass of the outer disks from the 90 $\mu$m fluxes of the stacked sources. The dust mass $M_{\mathrm{dust}}$ in an optically thin and isothermal disk is estimated as follows:

$$M_{\mathrm{dust}} = \frac{F_{\mathrm{disk},\nu} d^2}{\kappa_\nu B_\nu(T_{\mathrm{dust}})} \tag{10}$$

$$= 7.0 M_\oplus \left(\frac{F_{\mathrm{disk},\nu}}{100 \text{ mJy}}\right) \left(\frac{d}{150 \text{ pc}}\right)^2$$

$$\times \left(\frac{\kappa_\nu}{27.8 \text{ cm}^2\text{g}^{-1}}\right)^{-1} \left(\frac{B_\nu(T)}{B_\nu(20 \text{ K})}\right)^{-1}, \tag{11}$$

where $\nu$ denotes the frequency, $F_{\mathrm{disk},\nu}$ is the flux from the dust in the disk, $\kappa_\nu$ is the dust opacity, and $B_\nu(T)$ is the intensity of black body radiation at the dust temperature $T$. Here we assume $\kappa_\nu = 10(\nu/1200 \text{ GHz}) \text{ cm}^2\text{g}^{-1}$ (Cieza 2015). We also assume the characteristic temperature $T = 20$ K, at which the simplified isothermal model can reproduce results of a detailed SED-fit disk model in submillimeter wavelengths (Andrews & Williams 2007; Cieza 2015). We fix the distance $d$ at 150 pc, which is the average distance of the objects. The estimated dust masses are listed in table 4. Similar to the flux, the dust masses decrease with age. Note that the estimated dust mass also largely depends on the assumed dust temperature $T$. For example, when the assumed $T$ changes from 20 to 30 K, the estimated dust mass $M_{\rm dust}$ decreases by $\sim 1/8$.

Then, we estimate the age dependence of the dust mass in the outer disks. We can estimate the age dependence of the average dust mass based on that of the 90 $\mu$m fluxes as follows:

$$M_{\rm dust} = 60^{+20}_{-30} M_\oplus \exp\left(-\frac{t}{1.4^{+0.6}_{-0.5}\,{\rm Myr}}\right). \quad (12)$$

The decay timescale of the dust mass is the same as that of the 90 $\mu$m flux due to the assumption of their proportional relation. The uncertainties of the parameters are derived by the error propagation from the uncertainties of the dust masses and the typical ages.

Figure 9 shows the comparison of our results with those of other observations of disks at (sub)mm wavelengths. The flux in the (sub)mm wavelengths and that in the 90 $\mu$m band mainly come from the outer disks. Based on the (sub)mm fluxes, we estimate dust masses by equation (10). In the young and mid-age groups, the dust mass estimated from our result in the 90 $\mu$m band is lower than those estimated from observations at (sub)mm wavelengths. There are two possible interpretations of this difference of the dust masses. One interpretation is that only disks massive enough to be detected at (sub)mm wavelength are plotted on this graph, whereas our results show the typical behavior based on the stacking analysis. Our stacked images include faint objects whose fluxes are undetected with AKARI. Thus, through our stacking analysis, we can discuss dust-dissipation timescales in the outer disks containing faint objects. Another interpretation is that (sub)mm fluxes probe the dust in the outermost regions, which are extended farther than the regions emitting mainly far-IR fluxes. The outermost regions ($\sim 100$ au) probably contain more dust grains than the regions emitting far-IR fluxes ($\sim 10$ au).

To summarize the above discussion, we estimate the age dependence of the dust mass in the outer disks, from which the 90 $\mu$m flux mainly comes, and evaluate the decay timescale of the dust mass ($\tau_{\rm dust,outer} = 1.4^{+0.6}_{-0.5}$ Myr).

4.2 Comparison of the decay timescale at different wavelengths

To discuss the evolution of the different disk regions, we discuss the dust dissipation timescales at different disk regions based on the stacking results at different wavelengths.

First, we estimate the decay timescale of the intermediate disks emitting the mid-IR light. As with the case of the 90 $\mu$m flux, the decay timescale of the intermediate disk is assumed to be the same as that of the mid-IR flux. From the decay timescale of the $F_{\text{disk},22\mu\text{m}}$, the dust dissipation timescale of the intermediate disk is estimated as $\tau_{\text{dust,med}} = 1.38 \pm 0.05$ Myr.

**Table 4.** Estimated dust masses in each age group.

| group | $N_{\rm stack}$ | age [Myr] | $M_{\rm dust}[M_\oplus]^\dagger$ |
|---|---|---|---|
| young | 147 | $1.23^{+0.03}_{-0.04}$ | $23 \pm 4$ |
| mid-age | 769 | $4.44^{+0.03}_{-0.04}$ | $2.3 \pm 1.6$ |
| old | 3,867 | $11.96^{+0.05}_{-0.04}$ | $< 2.1\,(3\sigma)$ |

$^\dagger$ Disk dust masses estimated from the 90 $\mu$m disk fluxes. The errors are the $1\sigma$ uncertainties.

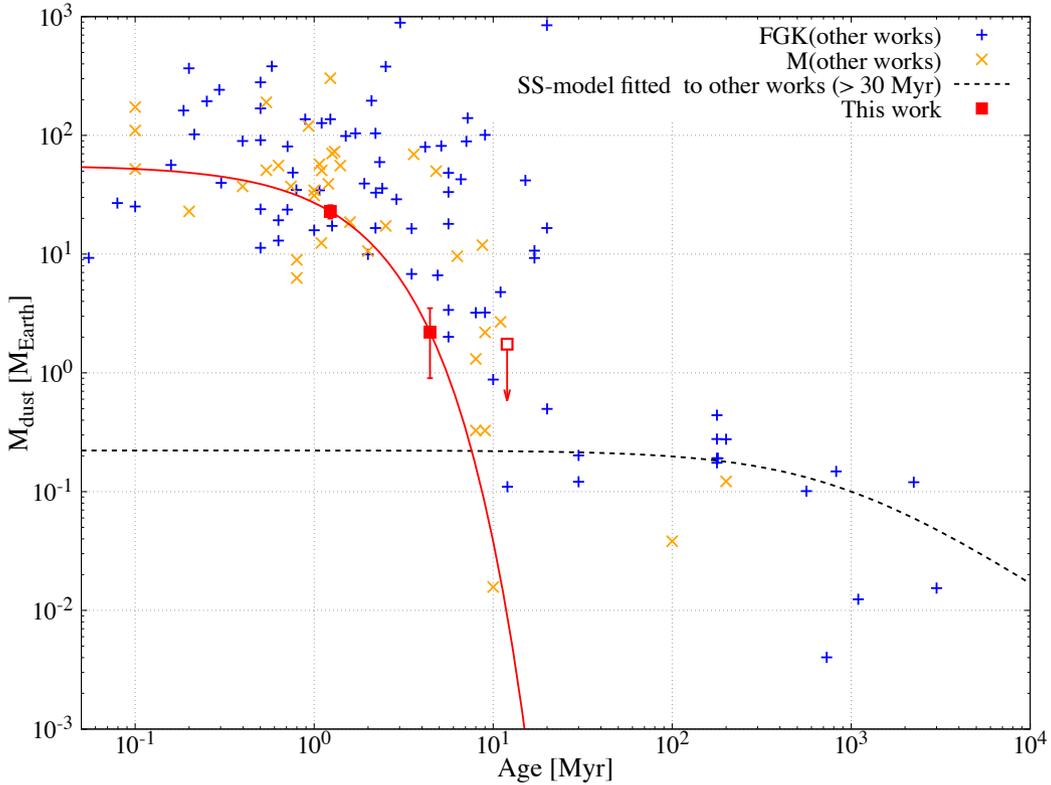

**Fig. 9.** Comparison of our estimated dust masses in disks with those derived from observed (sub)millimeter fluxes in the literature. The red squares show the estimated dust mass from the 90 $\mu$m flux in this work and the solid red line is the exponential-fitting line. Open squares show the $3\sigma$ upper limit of the $M_{\rm dust}$. Plus and cross symbols represent the FGK and M stars, respectively, obtained from previous (sub)mm observations: Jewitt & C. (1994); Osterloh & Beckwith (1995); Sylvester et al. (2001); Nuernberger et al. (1997); Nuernberger et al. (1998); Mannings & Sargent (1997); Mannings & Sargent (2000); Holland et al. (1998); Holland et al. (2003); Greaves et al. (2005); Williams & Andrews (2006); Lin et al. (2006); Matthews et al. (2007); Cieza et al. (2008); Roccatagliata et al. (2009); Ricci et al. (2010b); Ricci et al. (2010a); Nilsson et al. (2010). Only the detected objects in (sub)mm works are plotted. The black dashed line is fitted to the steady-state collisional evolution model of FGKM stars older than 30 Myr (see subsection 4.3).
(Color online)

From the current discussion and that in subsection 4.1, the dissipation time difference between the intermediate and outer disks is estimated as $\Delta\tau = \tau_{\text{dust,outer}} - \tau_{\text{dust,med}} = 0.0^{+0.6}_{-0.5}$ Myr. This time difference is not significantly different from zero. Thus, this result implies that dust in the intermediate and outer disks dissipates on almost the same timescale. This feature is consistent with the "two-timescale" disk evolution scenario, where the entire disk dissipates on almost the same timescale once the disk accretion stops (Williams & Cieza 2011).

The uncertainty of the $\Delta\tau$ is dominated by the photometric uncertainty of the 90 $\mu$m flux. To estimate the time difference at higher precision, we need to improve the photometric uncertainty of the far-IR flux.

Next, we compare our result with a previous study by Ribas et al. (2014, 2015). They analyzed the decay timescales of the ratio of YSOs having IR excesses at near-IR and mid-IR wavelengths. They derived the decay timescales of $\tau = 2.7 \pm 0.7$, $3.3 \pm 0.6$, and $5.3 \pm 0.9$ Myr at 3.4–4.6, 8–12, and 22–24 $\mu$m wavelengths, respectively. The decay timescales derived by us are 2–4 Myr shorter than that derived by them. This difference is considered to be due to differences in derivation methods. We derive the decay timescale of the disk fluxes. On the other hand, Ribas et al. (2014) derived the decay timescale of the ratio of YSOs having IR excesses, regardless of the value of IR excess. Since the disappearance of the IR excess occurs after the attenuation of the IR fluxes, the decay timescale of the disk flux derived by us is shorter than that of the ratio of YSOs having IR excesses derived by them. In addition, the decay timescales derived by us are more reliable than those derived by Ribas et al. (2014) in discussions on the decay timescale of the dust mass since the fluxes directly depend on the dust masses. Also, Ribas et al. (2014) showed that the decay timescale at longer wavelengths is longer. This is compatible with inside-out dust disk clearing among inner–intermediate disk range. Our results show that there is no significant difference between the decay timescales in the 12, 22, and 90 $\mu$m bands. Our results indicate that dust dissipates on almost the same timescale among the intermediate–outer disk regions and whether the dissipation is in an inside-out manner or not is unclear.

4.3 Comparison with debris disks

Here, we compare our results with a predicted dissipation timescale of dust disks, called debris disks, around main-sequence stars. Primordial dust is defined as dust originating in the interstellar medium.

The primordial dust remains small without growing into planetesimals and is considered to be dominant in young protoplanetary disks (Williams & Cieza 2011). On the other hand, secondary dust is defined as the dust formed by the collisional destruction of planetesimals (e.g., Wyatt 2008) and is dominant in debris disks. As a disk evolves from the protoplanetary disks to the debris disk, its dominant dust is considered to shift from primordial to secondary dust (e.g., Williams & Cieza 2011). However, the time of this transition is not clear. Revealing the time at which the dominant dust shifts from primordial to secondary would reveal the scenario of dust evolution and planet formation.

First, we compare our result with the dissipation timescale of the primordial dust. Observationally, the gas dissipation time in the outer disks has not been quantitatively assessed. Considering that few WTTSs in 1–10 Myr are detected in $^{12}$CO(2–1) line (Hardy et al. 2015), we assume that the gas dissipation timescale in the outer disks is <10 Myr. Once the gas has cleared, the dust motion is controlled by the radiation pressure to the gravitational-force ratio (Krivov 2010). Small dust dissipates by radiation pressure in approximately $10^{-3}$ Myr. Relatively large dust falls into the central star by the Poynting-Robertson (P-R) effect (Krivov 2010). The falling time $\tau_{\rm PR}$ under the P-R effect is

$$\tau_{\rm PR} = 8\left(\frac{r}{100\,{\rm au}}\right)^2 \left(\frac{M_*}{M_\odot}\right)^{-1} \left(\frac{\beta}{0.5}\right)^{-1} \,[{\rm Myr}]. \tag{13}$$

Here, $r$ is the distance of a dust grain from the central star, $M_*$ is the mass of the central star and $\beta$ is the ratio of radiation force to the gravitational force on the grain. Through the P-R effect and radiation pressure after gas clearing, the primordial dust in the outer disk dissipates over $1-10$ Myr, if we assume disks with the radii of $\sim 100$ au. The dissipation timescale of the outer disks in the current work ($1.4^{+0.6}_{-0.5}$ Myr) is within the predicted range of the dissipation timescale of primordial dust.

Next, we compare our result with the dissipation timescale of secondary dust. Because the secondary dust is formed by destructive events, dust production and dust dissipation occur simultaneously. Thus, the effective dissipation timescale of secondary dust is longer than that of the primordial one. The secondary dust dissipation can be described by a steady-state (SS) collisional evolutionary model (Wyatt et al. 2007). In the SS model, the disk dust mass $M_{\rm dust,SS}$ evolves in time ($t$) as follows:

$$M_{\rm dust,SS}(t) = \frac{M_{\rm dust,SS}(t=0)}{1 + \frac{t}{\tau_{\rm coll}(0)}}. \tag{14}$$

Here, $\tau_{\rm coll}(0)$ is the typical collision time of planetesimals at $t=0$ and $M_{\rm dust,SS}(t=0)$ is the initial dust mass. The evolutionary model is fitted to FGKM stars older than 30 Myr in previous studies plotted in figure 9 (Roccatagliata et al. 2009; Nilsson et al. 2010; Williams & Andrews 2006; Matthews et al. 2007) since stars older than 30 Myr are assumed to be surrounded not by protoplanetary disks but by debris disks. We perform this fitting on logarithmic scales to reduce the bias effect due to extremely bright outliers. As our fitting result, the free parameters are $M_{\rm dust,SS}(t=0) = 10^{-0.6\pm0.3} M_\oplus$ and $\tau_{\rm coll}(0) = 10^{2.4\pm0.5}$ Myr. The fitted collisional timescale $\tau_{\rm coll}(0)$ is within the range of the derived value by the previous studies ($10^2$–$10^3$ Myr in the outer disks; e.g., Kojima et al. in preparation.;Su et al. 2006). On the other hand, as discussed in subsection 4.1, the dust dissipation timescale in the outer disks is estimated as $1.4^{+0.6}_{-0.5}$ Myr. The $\tau_{\rm coll}$ is more than several tens of times longer than the dissipation timescale derived by us. Therefore, this comparison suggests that the secondary dust is not a major component in the disks of the young and mid-age groups ($\lesssim 6$ Myr).

From the above discussions, our work suggests that the major dust component in the outer disks is primordial dust, and the dissipation timescale is $1.4^{+0.6}_{-0.5}$ Myr.

Then, we discuss the transition time of the major dust component from primordial to secondary dust. As shown in figure 9, the black line of the fitted SS model and the red line of our work intersect at $\sim 8$ Myr. This result suggests that the dominant dust component changes from primordial to secondary at $\sim 8$ Myr.

We compare our results with the discussion by Panic et al. (2013) on the transition time from protoplanetary disks to debris disks. Panic et al. (2013) researched dust disk mass derived based on the objects detected in (sub)mm observation. They implied that the transition from protoplanetary disks to debris disks happens before 10–20 Myr, but this transition timescale has been unclear quantitatively since they discuss only detected objects. On the other hand, we discuss the decay timescale of the fluxes and the transition timescale from primordial to secondary dust, including non-detected objects in the 90 $\mu$m band. Since dust in protoplanetary disks is mainly primordial and that in debris disks is secondary, our result quantitatively suggests that the typical transition time from protoplanetary disks to debris disks is $\sim 8$ Myr.

## 5 Summary

To study the dissipation process in intermediate and outer protoplanetary disks, we stacked 12, 22, and 90 $\mu$m images of TTSs from WISE and AKARI all-sky images, and studied the decay time of the mid-IR and far-IR fluxes from disks. The main results are summarized below:

- The average 90 $\mu$m disk fluxes in young, middle-aged, and old disks decay over a timescale of $1.4^{+0.6}_{-0.5}$ Myr.
- The average 12 and 22 $\mu$m disk fluxes in young, middle-aged, and old disks decay over timescales of $1.4 \pm 0.2$ Myr and $1.38 \pm 0.05$ Myr, respectively.

Our results were further analyzed as follows:

- From the average 90 $\mu$m disk fluxes and with the assumption of isothermal thin disks, we estimate

the average dust disk mass around TTSs in each age group. Based on the decay time of the 90 $\mu$m disk fluxes, we estimate the dust-dissipation timescale in the outer disk as $\tau_{\text{dust,outer}} = 1.4^{+0.6}_{-0.5}$ Myr.

- Based on the decay times of the 12, 22, and 90 $\mu$m disk fluxes, the time difference between the dust dissipation times in the intermediate and outer disks is estimated as $\Delta\tau_{\text{dust}} = 0.0^{+0.6}_{-0.5}$ Myr. This implies that the intermediate and outer disks dissipate on almost the same timescale.

- We compare the obtained dust dissipation timescale with the predicted dissipation timescale of primordial and secondary dust. By this comparison, we conclude that the dust in the outer disks at $\lesssim 8$ Myr is primordial. Through comparison with the evolutionary model of debris disks, our result suggests that the transition timescale of the dominant dust from primordial to secondary is $\sim 8$ Myr. This timescale has not been quantitatively clear in past studies.

**Acknowledgments**

This research is based on observations with AKARI, a JAXA project with the participation of ESA. We are grateful to Professor T. Suzuki for his informative advice. We thank Mr. Yasuhiro Matsuki for the original code of the stacking analysis, which is the basis of our studies. This publication makes use of data products from the Wide-field Infrared Survey Explorer, which is a joint project of the University of California, Los Angeles, and the Jet Propulsion Laboratory/California Institute of Technology, and NEOWISE, which is a project of the Jet Propulsion Laboratory/California Institute of Technology. WISE and NEOWISE are funded by the National Aeronautics and Space Administration. This work has made use of data from the European Space Agency (ESA) mission *Gaia* (https://www.cosmos.esa.int/gaia), processed by the *Gaia* Data Processing and Analysis Consortium (DPAC, https://www.cosmos.esa.int/web/gaia/dpac/consortium). Funding for the DPAC has been provided by national institutions, in particular the institutions participating in the *Gaia* Multilateral Agreement. This publication makes use of data products from the Two Micron All Sky Survey, which is a joint project of the University of Massachusetts and the Infrared Processing and Analysis Center/California Institute of Technology, funded by the National Aeronautics and Space Administration and the National Science Foundation. The authors would like to thank Enago (www.enago.jp) for the English language review. H.M. appreciates the financial supports by Advanced Leading Graduate Course for Photon Science (ALPS) of the University of Tokyo.

**Appendix 1  The effect of the distance and color distributions**

In the stacking analysis, we do not apply the correction of individual samples' distances. Also, we do not consider the correction for the stellar mass. Here, we discuss the effect of the distance and the stellar mass distributions on the stacking analysis results.

Figure 10 shows the distance histogram of each age-group sample. The average distances of the young, mid-age, and old groups are 155, 147, and 145 pc, respectively, with standard deviations of 24, 23, and 26 pc, respectively. The standard errors are 2.0, 0.8, and 0.4 pc for the young, mid-age, and old groups, respectively. The difference in the average distances is larger than the three times standard errors.

From the average distance $d_{\text{average}}$ of an age group, we estimate the flux $F_{\text{corrected},\nu}$ of the

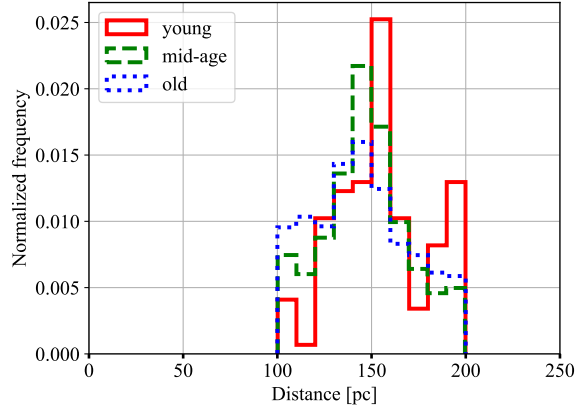

**Fig. 10.** The distance histogram of samples in each age group. The vertical axis shows the normalized frequency. The range of distance is limited to 100–200 pc in (a3) screening (see section 2.2.1). (Color online)

**Table 5.** The decay time of the disk flux with and without the distance correction. The decay time without and with the correction is denoted by $\tau_\nu$ and $\tau_{\text{corrected},\nu}$, respectively.

| $\lambda_{\text{eff}}$ [$\mu$m] | $\tau_\nu$ [Myr] | $\tau_{\text{corrected},\nu}$ [Myr] |
|---|---|---|
| 12 | $1.4 \pm 0.2$ | $1.3 \pm 0.2$ |
| 22 | $1.38 \pm 0.05$ | $1.32 \pm 0.05$ |
| 90 | $1.4^{+0.6}_{-0.5}$ | $1.3^{+0.6}_{-0.5}$ |

stacked source with the distance correction at 150 pc as follows:

$$F_{\text{corrected},\nu} = F_\nu \left(\frac{d_{\text{average}}}{150 \text{ pc}}\right)^2, \tag{A1}$$

where $F_\nu$ is the flux derived in section 3. The corrected fluxes are estimated as $F_{\text{corrected},\nu} = 1.07 F_\nu, 0.96 F_\nu, 0.94 F_\nu$ for the young, mid-age, and old groups, respectively.

Then, we discuss the effect of the distance correction on the decay time estimation. Since the distance correction factor is the same for all the wavelength bands in the same group, the correction factor of the disk flux is the same as that of the $F_\nu$. Based on the correction factor, we estimate the decay time $\tau_{\text{corrected}}$ of the disk flux with the distance correction. Table 5 shows the comparison between the decay time with and without the distance correction. The decay time with the correction is only $\sim 5\%$ shorter than that without the correction in all the wavelength bands. Therefore, we conclude that the distance bias between the groups does not affect our results and discussions.

Next, we discuss the bias of the stellar mass distribution. Since the evolution of low-mass PMS stars follows the Hayashi track (Hayahi 1961), the stellar mass can be estimated from the intrinsic

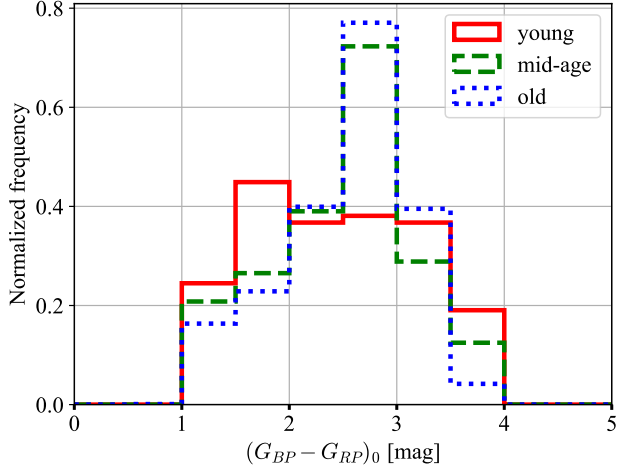

**Fig. 11.** The color $G_{\rm BP} - G_{\rm RP}$ histogram of samples in each age group. The vertical axis shows the normalized frequency. The range of the color is restricted to 0.994–3.8 mag (see sections 2.2.1 and 2.3). (Color online)

color. Here, we thus discuss the bias of the color distribution of each group. Figure 11 shows the histogram of the intrinsic color $(G_{\rm BP} - G_{\rm RP})_0$ of each age group. The average color of the young, mid-age, and old groups are $(G_{\rm BP} - G_{\rm RP})_0 =$ 2.45, 2.49, and 2.53 mag, respectively, with standard deviations of 0.7, 0.7, and 0.6 mag, respectively. The standard errors are 0.05, 0.03, and 0.01 mag for the young, mid-age, and old groups, respectively. Since the difference in the average colors is smaller than three times the standard error, we consider that the color bias is negligible.

To summarize, the distance and color distribution biases between age groups are so small that they do not affect our discussions.

## Appendix 2  Disk flux excess over fitted-curve in the 22 $\mu$m band of the old group

When comparing the decay time of the disk fluxes in the 22 $\mu$m bands for objects in the same age group (subsection 3.3), we exclude the point of the old group from the curve-fitting due to the non-detection of the 12 and 90 $\mu$m disk fluxes of the old group. As shown in figure 8, the 22 $\mu$m disk flux of the old group is higher than the fitted curve. Here, we discuss the possible reasons for these excesses.

First, the mass ratio of secondary dust in total dust might be larger in the disks of the old group than the young and mid-age groups. Many debris disks at the age of $\gtrsim$ 10 Myr have been observed (e.g., Wyatt 2008). As described in subsection 4.3, the effective dissipation timescale of secondary dust is longer than that of primordial dust. The presence of the young debris disks would raise the average flux from the old group above that expected from only the primordial dust dispersal.

Second, the evolutionary curve in old disks may be poorly fitted to exponential functions. The exponential evolution of fluxes is only an assumption. If the decay rate of the fluxes slows over time, the observational values may exceed the fitted curves.